\documentclass[11pt]{article}
\usepackage{graphicx}
\usepackage{cite}

\begin{document}

%
% Title, Authors, Affiliations
% ===================
%
\title{\Large\bf
Transverse single spin asymmetries at small x
 and the anomalous magnetic moment }

\author{Jian Zhou
 \\[0.3cm]
{\normalsize\it Institut f\"{u}r Theoretische Physik,Universit\"{a}t
Regensburg, Regensburg, Germany}}

\maketitle

\begin{abstract}
\noindent We show that in the Mclerran-Venugopalan model, an axial asymmetrical valence quark distributions in the
transverse plane of a transversely polarized proton can give rise to a spin dependent odderon.
Such polarized odderon is responsible for the transverse single spin asymmetries(SSAs) for jet production
in the backward region of pp collisions and open charm production in semi-inclusive DIS process.
\end{abstract}

\section{Introduction }

The exploration of transverse single spin asymmetries(SSAs)
in high energy scattering experiments has a long history, starting from the mid 1970¡¯s~\cite{Bunce:1976yb}.
The large size of the observed SSAs for single inclusive hadron production
 came as a big surprise and, a priori, posed a challenge
for QCD, as the naive parton model predicts the asymmetries are proportional to
the quark mass~\cite{Kane:1978nd,Ma:2008gm} and thus very small.
During the past few decades,  the remarkable theoretical progress was achieved
by going beyond the naive parton model and following mainly two approaches: one approach is based on transverse momentum dependent(TMD)
factorization~\cite{Sivers:1990fh, Collins:1992kk}
and the other on collinear twist-3
factorization~\cite{Efremov:1981sh,Qiu:1991pp,Ji:1992eu,Yuan:2009dw,Beppu:2010qn}.
In TMD factorization,  naive time reversal odd TMD distributions and fragmentation function,
known as the quark/gluon Sivers functions~\cite{Sivers:1990fh} and the Collins fragmentation function~\cite{Collins:1992kk}
 can account for the large SSAs, while in the collinear twist-3 approach,
the SSAs arise from twist-3 quark gluon correlator
so-called Efremov-Teryaev-Qiu-Sterman function (ETQS)~\cite{Efremov:1981sh,Qiu:1991pp}, tri-gluon
correlation functions~\cite{Ji:1992eu,Beppu:2010qn}, and twist-3 collinear fragmentation functions~\cite{Yuan:2009dw}.

Beyond TMD factorization and collinear twist-3 formalism,
some alternative mechanisms underlying the large SSAs have been proposed, such as
the soft coherent dynamics~\cite{Hoyer:2006hu}, QCD instanton mechanism\cite{Kochelev:1996pv,Qian:2011ya},
and QCD odderon interaction~\cite{Ahmedov:1999ne,Kovchegov:2012ga}.
The authors of papers~\cite{Ahmedov:1999ne} investigated the odderon's contribution
to SSAs in the context of a heavy fermion model. A later calculation formulated in the
saturation/color glass condensate(CGC) framework suggests that SSAs can be generated by the interaction of
spin-dependent light-cone wave function of the projectile with the target gluon field via C-odd odderon exchange~\cite{Kovchegov:2012ga}.
The odderon excitation considered in~\cite{Kovchegov:2012ga} comes from the unpolarized proton/nucleus and is
spin independent. In this paper, however,
 we focus on studying SSAs generated by a spin dependent odderon that comes from the transversely polarized proton.
Such polarized odderon is responsible for SSAs for jet production
in the backward region of polarized proton in pp collisions and open charm production in SIDIS process at small x.

In recent years, the interplay between spin physics and saturation physics is forming into an active field of research.
The early work includes the study of small x evolution of spin dependent structure function $g_1$~\cite{Bartels:1995iu}.
It was pointed out that the spin asymmetries could also be generated by the
 pomeron-odderon interference effect~\cite{oai:arXiv.org:hep-ph/0202231}.
SSAs at forward rapidity in pA collisions have been investigated in Refs.~\cite{Boer:2006rj,Kovchegov:2012ga}.
More recently, the quark/gluon Boer-Mulders distributions inside a large nucleus have been studied
in Refs.~\cite{Metz:2011wb}.

In this paper, we explore  SSA phenomena at small x and identify a spin dependent odderon as the main source of SSAs.
The paper is organized as follows. In Sec.2, we show that, in the MV model, non-vanishing contribution to the spin dependent
 odderon amplitude arises from the left-right asymmetrical color source distribution in the transverse plane of a transversely
polarized proton. In Sec.3, we present compact expressions for SSAs in jet production in the backward region of pp
collisions and open charm production in SIDIS at small x. Both asymmetries are generated by the polarized odderon.
 We summarize our paper in Sec.4.

\section{A spin dependent classical  odderon }
In perturbative QCD, the odderon is a color-singlet exchange
and can be formed by three gluons in a symmetric color state.
It has negative C-parity and therefore dominates the differences between particle-particle and particle-antiparticle scatterings at high energy.
The energy dependence of the odderon exchange is described by the BKP equation~\cite{Bartels:1980pe}.
Within the CGC formalism, one can identify the following operator as the dipole odderon operator~\cite{Hatta:2005as},
\begin{eqnarray}
\hat O(R_\perp,r_\perp)=\frac{1}{2i} \left [ \hat D(R_\perp,r_\perp)-\hat D(R_\perp, -r_\perp) \right ] \ ,
\end{eqnarray}
where
\begin{eqnarray}
\hat D(R_\perp, r_\perp)=\frac{1}{N_c}
{\rm Tr} \left [ U(R_\perp+\frac{r_\perp}{2}) U^\dag(R_\perp-\frac{r_\perp}{2} ) \right ] \ ,
\end{eqnarray}
with the Wilson line being defined as,
\begin{eqnarray}
U(x_\perp)= {\rm P} e^{ig \int_{-\infty}^{+\infty} dx^- A_+(x^-, \ x_\perp)}  \ .
\end{eqnarray}
The small x evolution equation of this odderon operator was constructed using the dipole model\cite{Kovchegov:2003dm}
 and the general JIMWLK equation~\cite{Hatta:2005as}.
In the low parton densities region, the BLV solution~\cite{Bartels:1999yt} to the BKP equation can be recovered from both
the dipole model calculation and the JIMWLK equation with C-odd initial conditions.

The odderon is absent in the original MV model~\cite{McLerran:1993ni} in which the distribution of  large x color source is assumed to take
a Gaussian form. However,  it has been shown that a classical odderon can be generated by
 an additional cubic term  in the modified weight function $W[\rho]$,
 which is given by~\cite{Jeon:2004rk},
\begin{eqnarray}
W[\rho]={\rm exp} \left \{
-\int d^2 x_\perp \left [ \frac{\rho_a(x_\perp) \rho_a(x_\perp)}{2\mu (x_\perp)}
- \frac{g d_{abc} \rho^a(x_\perp) \rho^b(x_\perp)\rho^c(x_\perp) }{4N_c\mu^2(x_\perp)} \right ]
\right \} \ ,
\end{eqnarray}
where $\mu(x_\perp)$is the density of color sources per unit transverse area and related to the valence quark distribution
in the transverse plane: $\int dx_q f_q(x_q, x_\perp)=6  \mu(x_\perp)/g^2$ with $x_q$ being the longitudinal momentum fraction
carried by valence quark;
and $d^{bca}$ is the symmetric structure constant of the color SU(3) group.
Using the above weight function to compute the expectation value of the odderon operator, one obtains~\cite{Jeon:2004rk}
\begin{eqnarray}
&& \int d^2 R_\perp \theta(R_0-|R_\perp|) < \hat O(R_\perp, r_\perp) >
\nonumber\\ &&
=c_0 \alpha_s^3 \int d^2 R_\perp \theta(R_0-|R_\perp|)
\int d^2 z_\perp  {\rm ln}^3 \frac{|R_\perp+r_\perp/2-z_\perp|}{|R_\perp-r_\perp/2-z_\perp|}
e^{-\frac{1}{4}r_\perp^2 Q_s^2} \frac{1}{3} \int d x_q f_q(x_q,z_\perp)
\label{odderon0}
\end{eqnarray}
where $R_0$ is the radius of proton
and the transverse center of  parton longitudinal momentum  is chosen to be as the origin.
The color coefficient $c_0$ is defined as $c_0=\frac{(N_c^2-1)(N_c^2-4)}{4 N_c^3} $.
$Q_s^2 = \alpha_s C_F \mu(R_\perp) {\rm ln} \frac{1}{r_\perp^2 \Lambda_{QCD}^2}$ is the quark saturation momentum.
Here we insert a theta function
 because we have assumed that the dipole must hit the proton directly in order to be able to interact with
quarks inside of it.

To proceed further, we first neglect the dependence of $Q_s^2$ on $R_\perp$ and $r_\perp$. In order to
integrate out $R_\perp$, we use a mathematical trick introduced in~\cite{Kovchegov:2012ga}. One notices that
\begin{eqnarray}
\int d^2 R_\perp{\rm ln}^3 \frac{|R_\perp+r_\perp/2-z_\perp|}{|R_\perp-r_\perp/2-z_\perp|}=0
\end{eqnarray}
if the integration carries over the whole transverse plane.  This result implies,
\begin{eqnarray}
\int d^2 R_\perp{\rm ln}^3 \frac{|R_\perp+r_\perp/2-z_\perp|}{|R_\perp-r_\perp/2-z_\perp|} \theta(R_0-|R_\perp|)
=-\int d^2 R_\perp{\rm ln}^3 \frac{|R_\perp+r_\perp/2-z_\perp|}{|R_\perp-r_\perp/2-z_\perp|}\theta(|R_\perp|-R_0) \ .
\end{eqnarray}
The fact that $|r_\perp|$ is much smaller than $R_0$ for a perturbative dipole  allows us to expand
the integrand on the right-hand-side of the above equation in powers
of $|r_\perp|/|2R_\perp|$ as well as $|z_\perp|/|R_\perp|$. To the first non-trivial order, one has,
\begin{eqnarray}
\int d^2 R_\perp{\rm ln}^3 \frac{|R_\perp+r_\perp/2-z_\perp|}{|R_\perp-r_\perp/2-z_\perp|} \theta(R_0-|R_\perp|)
\approx - \frac{3 \pi}{4 R_0^2} r_\perp^2 (r_\perp \cdot z_\perp)
\label{mathtrick} \ .
\end{eqnarray}
Substituting Eq.~\ref{mathtrick} back into Eq.~\ref{odderon0}, we obtain,
\begin{eqnarray}
\int d^2 R_\perp  \theta(R_0-|R_\perp|) \langle \hat O(R_\perp, r_\perp) \rangle
\approx-\frac{ c_0 \alpha_s^3  \pi}{4 R_0^2} r_\perp^2 e^{-\frac{1}{4}r_\perp^2 Q_s^2}
\int dx_q d^2 z_\perp  (r_\perp \cdot z_\perp)
 f_q(x_q, z_\perp) \ .
 \label{odderon1}
\end{eqnarray}
For a transversely polarized proton, impact parameter dependent valence quark distribution can be parameterized as~\cite{Burkardt:2000za},
\begin{eqnarray}
 f_q(x_q,z_\perp)=  \sum_{u,d} \left \{ {\cal H} (x_q,z_\perp^2)- \frac{1}{2M} \epsilon_{\perp}^{ij}S_{\perp i}
 \frac{\partial{\cal E}(x_q,z_\perp^2)}{ \partial z_{\perp}^j} \right \}
 \label{GPDs} \ ,
\end{eqnarray}
where $S_\perp$ is the proton transverse spin vector and $M$ is the proton mass.
The generalized parton distributions(GPDs) ${\cal H}(x_q,z_\perp^2)$ and ${\cal E}(x_q,z_\perp^2)$ are the
Fourier transformed GPDs $H$ and $E$ with zero skewedness, respectively.
Inserting Eq.~\ref{GPDs} into Eq.~\ref{odderon1}, one immediately obtains
\begin{eqnarray}
\int d^2 R_\perp \theta(R_0-|R_\perp|) < \hat O(R_\perp, r_\perp) >
&=&- \frac{ c_0 \alpha_s^3 \pi  }{8 M R_0^2} e^{-\frac{1}{4}r_\perp^2 Q_s^2}
r_\perp^2 \epsilon_{\perp}^{ij}S_{\perp i}  r_{\perp j}
\int d x_q d^2z_\perp  \sum_{u,d} {\cal E}(x_q,z_\perp^2) \
\nonumber\\
&=&- \frac{ c_0 \alpha_s^3 \pi  }{8 M R_0^2} e^{-\frac{1}{4}r_\perp^2 Q_s^2}
r_\perp^2 \epsilon_{\perp}^{ij}S_{\perp i}  r_{\perp j}
\left ( \kappa_p^u + \kappa_p^d   \right ) \ ,
\label{odderon2}
\end{eqnarray}
where $ \kappa_p^u $ and $ \kappa_p^d $ are the contributions from up and down quarks  to the anomalous magnetic moment
of the proton, respectively. An earlier attempt to connect SSA phenomena to GPD E has been made in paper~\cite{Burkardt:2002ks}.

A few comments are in order on the above analytic result.
\begin{itemize}
\item
First, the odderon exchange under our consideration is clearly  spin dependent.
\item
The polarized odderon originates from the transverse distortion of the
impact parameter dependent PDF inside a transversely polarized proton.
Such transverse distortion has been clearly seen in a lattice QCD calculation~\cite{Gockeler:2006zu}.
\item
Given the nucleon's magnetic moment  $ \kappa_p=1.793$ and $ \kappa_n=-1.913$,
 $ \kappa_p^u $ and $ \kappa_p^d $ can be roughly estimated as $ \kappa_p^u=1.673 $ and $ \kappa_p^d =-2.033 $
 by using the isospin symmetry.
 Obviously, the contributions from u and d quarks to the polarized odderon   largely cancel out.
\end{itemize}

We conclude this section by  making a final remark on our result.
The MV model is expected to work better for a large nucleus.
However, its application to a proton target turns out to be quite successful phenomenologically~\cite{Albacete:2010sy}.
Therefore, qualitatively speaking, our analysis presented here might be relevant in phenomenological studies as well.

\section{Observables }
As discussed in the introduction, the spin dependent odderon is responsible for the transverse single spin asymmetry for
jet production in the backward region of the polarized proton in pp collisions.  The leading order result for jet production
was first derived in Ref.~\cite{Dumitru:2002qt}. Recently, the NLO correction to the cross section has also been calculated
in papers~\cite{Chirilli:2011km}.

At leading order, for the quark-initiated subprocess, the cross section reads
\begin{eqnarray}
\frac{d \sigma^{pA \longrightarrow qX}}{d^2 k_\perp dY}
= \sum_f x q_f(x)
\int \frac{d^2r_\perp }{(2\pi)^2} e^{-i  k_\perp \cdot r_\perp} \int d^2 R_\perp
\langle \hat D(R_\perp,r_\perp) \rangle_{x_g} \ ,
\end{eqnarray}
where $x=\frac{|k_\perp| }{\sqrt{s}}e^{-Y}$ and $x_g=\frac{|k_\perp| }{\sqrt{s}}e^{Y}$, with $Y$ being the rapidity.
$q_f(x)$ is the normal integrated quark distribution from the  unpolarized proton.
Note that we have neglected elastic scattering contribution to the cross section in the above expression.
The dipole S-matrix can be decomposed into the even and odd pieces under the exchange of the transverse coordinates
\begin{eqnarray}
\hat D(R_\perp,r_\perp)=\hat S(R_\perp,r_\perp)+i \hat O(R_\perp,r_\perp) \ ,
\end{eqnarray}
with the symmetric part being defined as
\begin{eqnarray}
\hat S(R_\perp,r_\perp)=\frac{1}{2} \left [ \hat D(R_\perp,r_\perp)+\hat D(R_\perp,-r_\perp) \right ] \ ,
\end{eqnarray}
The cross section then can be re-expressed as
\begin{eqnarray}
\frac{d \sigma^{pA \longrightarrow qX}}{d^2 k_\perp dY}&=&\sum_f x q_f(x)
\int \frac{d^2r_\perp }{(2\pi)^2} e^{-i  k_\perp \cdot r_\perp} \int d^2 R_\perp
\langle \hat S(R_\perp, r_\perp)+i \hat O(R_\perp, r_\perp) \rangle_{x_g}
\nonumber\\ &=&
\sum_f x q_f(x)\left \{  F_{x_g}(k_\perp^2)+\frac{1}{M}
 \epsilon_{\perp}^{ij}S_{\perp i}  k_{\perp j} O_{1T, x_g}^\perp(k_\perp^2) \right \} \ .
\end{eqnarray}
Here we introduce a spin dependent odderon in momentum space: $O_{1T, x_g}^\perp(k_\perp^2)$.
 To some extent, $O_{1T, x_g}^\perp(k_\perp^2)$ can be considered as a C-odd partner of the gluon Sivers function.
In the MV model, the unpolarized gluon distribution is given by,
\begin{eqnarray}
  F_{x_g}(k_\perp^2)=\pi R_0^2 \int \frac{d^2r_\perp }{(2\pi)^2} e^{-i  k_\perp \cdot r_\perp}
e^{-\frac{1}{4} r_\perp^2 Q_{s}^2} \ .
\label{unpolarized}
\end{eqnarray}
Using Eq.~\ref{odderon2}, it is easy to derive,
\begin{eqnarray}
O_{1T, x_g}^\perp(k_\perp^2)=
\frac{ -c_0 \alpha_s^3 \left ( \kappa_p^u + \kappa_p^d   \right )  }{4 R_0^4} \left [
\frac{\partial}{\partial k_\perp^2 } \frac{\partial}{\partial k_\perp^i} \frac{\partial}{\partial k_{\perp i}}
F_{x_g}(k_\perp^2) \right ] \ .
\label{polarized}
\end{eqnarray}
A few comments are in order on the above analytic result.
\begin{itemize}
\item
We note that  $\int d^2 k_\perp k^2_\perp O_{1T, x_g}^\perp(k_\perp^2)=0$.
This relation implies that $O_{1T, x_g}^\perp(k_\perp^2)$ has a node in $k_\perp^2$,
 and the $k_\perp^i$ weighted cross section $\int d^2k_\perp <k_\perp^i d \sigma >$ is zero.
\item
The single spin asymmetry is determined by the ratio $k_\perp O_{1T, x_g}^\perp/F_{x_g}$.
From Eq.~[\ref{unpolarized}] and Eq.~[\ref{polarized}], one finds that
this ratio scales as $k_\perp$ at low transverse momentum, while it scales as $1/k_\perp^3$ at high transverse momentum.
\item
The ratio $k_\perp O_{1T, x_g}^\perp/F_{x_g}$ should drop with decreasing $x_g$ as the power of $(x_g)^{0.3}$ since the leading high energy odderon intercept
is equal to 1 according to the BLV solution.
\end{itemize}
For the gluon initiated channel, the cross section reads
\begin{eqnarray}
\frac{d \sigma^{pA \longrightarrow gX}}{d^2 k_\perp dY}
=  x g(x)
\int \frac{d^2r_\perp }{(2\pi)^2} e^{-i  k_\perp \cdot r_\perp} \int d^2 R_\perp
\langle \hat { \tilde D}(R_\perp,r_\perp) \rangle_{x_g} \ ,
\end{eqnarray}
where $g(x)$ is the normal integrated gluon distribution from the unpolarized proton,
 and $\hat { \tilde D}(r_\perp)$ is given by,
\begin{eqnarray}
\hat {\tilde D}(R_\perp, r_\perp)=\frac{1}{N_c^2-1}
{\rm Tr} \left [ \tilde U(R_\perp+\frac{r_\perp}{2}) \tilde U^\dag(R_\perp-\frac{r_\perp}{2} ) \right ]
\end{eqnarray}
with $\tilde U$ being the Wilson line in the adjoint representation.
In the large $N_c$ limit, the cross section is approximated as,
\begin{eqnarray}
 \frac{d \sigma^{pA \longrightarrow gX}}{d^2 k_\perp dY}
\approx
x g(x)
\int \frac{d^2r_\perp }{(2\pi)^2} e^{-i  k_\perp \cdot r_\perp} \int d^2 R_\perp
{ \Big \langle} \left  \{[ \hat S(R_\perp, r_\perp)]^2+[ \hat O(R_\perp, r_\perp)]^2 \right  \} {\Big \rangle}_{x_g}
 \end{eqnarray}
where all antisymmetric interference terms $\hat S(R_\perp, r_\perp)\hat O(R_\perp, r_\perp)$ completely cancel out.
We are only left with the symmetric terms that do not contribute to the spin asymmetry.
Since SSA vanishes in the gluon initiated jet production process, one might expect that the spin asymmetry
rises with the increasing $|x_F|$ in the backward region of pp collisions.

Recently, the single spin asymmetry in inclusive jet production has been measured in both forward and backward regions
at the AnDY experiment at RHIC~\cite{Nogach:2012sh}.
There is at least one experimental data point in the backward region which is inconsistent
with zero within error bar. The possible two sources of the spin asymmetry in the backward region are the
polarized odderon and the gluon Sivers function. However, we have shown that the gluon Sivers function
dies out very quickly with decreasing $x_g$~\cite{Schafer:2013opa}.
Therefore, this measurement likely indicates that the polarized odderon indeed exists.

Let us now turn to discuss the SSA in open charm production in SIDIS process.
The differential cross section for this process has been calculated in the dipole model~\cite{Mueller:1999wm}
 and in the CGC formalism~\cite{McLerran:1998nk}. The next to leading order correction to this process
 is also available in Refs.~\cite{Balitsky:2010ze,Beuf:2011xd}.
At leading order, the cross section in  momentum space reads,
\begin{eqnarray}
\frac{d \sigma}{dx_B dz  dQ^2 dy  d^2 l_{\perp}}&=&\frac{
\alpha_{em}^2 e_c^2}{2 \pi^4 x_B Q^2} \left [ 1-y+\frac{y^2}{2}\right  ] \left [ z^2+(1-z)^2 \right ]
\nonumber\\&& \times
\int \frac{d^2
p_\perp}{(2\pi)^2} \frac{d^2 k_{\perp}}{(2\pi)^2} \frac{d^2
k_{\perp}'}{(2\pi)^2}
\frac{(l_\perp-k_{\perp})\cdot
(l_\perp-k_\perp')}{[\rho+(l_\perp-k_\perp)^2][\rho+(l_\perp-k_\perp')^2]}
\nonumber\\&&  \times { \Big \langle} {\rm Tr} \Big \{ \left [
U(k_\perp)U^\dag(k_\perp-p_\perp)-(2\pi)^4\delta^2(k_\perp)\delta^2(k_\perp-p_\perp) \right ]
\nonumber\\&&   \ \ \ \ \ \ \ \
  \times \left [
U(k_\perp'-p_\perp) U^\dag(k_\perp')-(2\pi)^4\delta^2(k_\perp')\delta^2(k_\perp'-p_\perp) \right ] \Big \}
{\Big \rangle}_{x_g}
\end{eqnarray}
Here the common kinematical variables in SIDIS process
 are defined as $Q^2=-q \cdot q$, $x_B=Q^2/2P\cdot q$, $y=q\cdot P/P_e \cdot P$
 and $z=l \cdot P/P\cdot q$ where $l$, $P_e$, $P$, and $q$  are momenta for produced charm quark,
 incoming lepton and proton, and virtual photon, respectively.
$\rho$ is defined as $\rho=z(1-z)Q^2$.
$U(k_\perp)$ is the Fourier transform of $U(x_\perp)$.
For simplicity, we have neglected  charm quark mass and only taken into account the
transverse polarized virtual photon contribution to the differential cross section.
 The above formula can be re-organized and expressed in a more compact form:
\begin{eqnarray}
\frac{d \sigma}{dx_B dz  dQ^2 dy  d^2 l_{\perp}}&=&\frac{
\alpha_{em}^2 e_c^2}{2 \pi^4 x_B Q^2}
\int \frac{d^2 k_{\perp}}{(2\pi)^2}
 H(k_\perp,l_\perp,Q^2)
{\Big \langle} {\rm Tr} \left [
U(k_\perp)U^\dag(k_\perp) \right ] {\Big \rangle}_{x_g}
\nonumber\\&=&  \frac{
\alpha_{em}^2 e_c^2N_c}{2 \pi^4 x_B Q^2}
 \int d^2 k_{\perp} H(k_\perp,l_\perp,Q^2)
\left [  F_{x_g}(k_\perp^2)+
 \frac{1}{M} \epsilon_{\perp}^{ ij}S_{\perp i}  k_{\perp j} O_{1T, x_g}^\perp(k_\perp^2) \right ] \ ,
\end{eqnarray}
where the first term recovers the known unpolarized differential cross section,
 whereas the second term is the spin dependent contribution.
For anti-charm quark production, $U(k_\perp)U^\dag(k_\perp)$ that appears in the first line of the above equation should be replaced with
$U^\dag(k_\perp)U(k_\perp)$, leading to the exactly opposite SSA as compared to that in charm quark production.
The hard part $H(k_\perp,l_\perp,Q^2)$ is given by,
\begin{eqnarray}
H(k_\perp,l_\perp,Q^2)=\left [ 1-y+\frac{y^2}{2} \right ] \left [ z^2+(1-z)^2 \right ]
\left [ \frac{l_\perp-k_\perp}{\rho+(l_\perp-k_\perp)^2}-
\frac{l_\perp}{\rho+l_\perp^2} \right ]^2  \ .
\end{eqnarray}

The SSA in open charm production in SIDIS process has also been
calculated using the collinear twist-3 approach~\cite{Beppu:2010qn}(for earlier work, see~\cite{Kang:2008qh}).
In the framework of the collinear factorization, a C-odd tri-gluon
correlation which gives rise to SSA is defined as~\cite{Ji:1992eu,Beppu:2010qn},
\begin{eqnarray}
O^{\alpha \beta \gamma}(x_1,x_2)=
-gi^3\int \frac{dy^- dz^-}{(2\pi)^2 P^+} e^{iy^- x_1P^+}e^{iz^-(x_2- x_1)P^+}
\langle pS| d^{bca} F_b^{\beta+}(0)F_c^{\gamma +}(z^-)F_a^{\alpha+}(y^-)|pS \rangle
\label{tri-gluon}
\end{eqnarray}
where we regard all the free Lorentz indices $\alpha$, $\beta$ and $\gamma$ to be transverse in three dimension.
One can also define a C-even tri-gluon correlation $N^{\alpha \beta \gamma}(x_1,x_2)$ by replacing
$d^{bca}$ with the anti-symmetric tensor $if^{bca}$ in the above equation~\cite{Ji:1992eu,Beppu:2010qn}.
Both the C-even and C-odd tri-gluon correlations contribute to SSAs. However, only the C-odd tri-gluon correlation
is the relevant one at small x as shown in~\cite{Schafer:2013opa}.

It is known that the $k_\perp$ moment of the gluon Sivers function can be related to the gluonic pole C-even
tri-gluon correlation $N^{\alpha \beta \gamma}(x_g,x_g)$. A similar relation between the $k_\perp$ moment of the
polarized odderon and the C-odd tri-gluon correlation can be established.
At small x, exponentials that appear in Eq.~\ref{tri-gluon} can be approximated as
 $e^{iy^- x_1P^+} \approx1 $ and $e^{iz^-(x_2- x_1)P^+}\approx 1$.
In this approximation, one has
\begin{eqnarray}
 \int d^2 k_\perp k_\perp^\alpha k_\perp^\beta k_\perp^\gamma
\frac{1}{M} \epsilon_{\perp}^{ij}S_{\perp i}  k_{\perp j} O_{1T, x_g}^\perp(k_\perp^2)
= \frac{-i g^2 \pi^2}{2 N_c }O^{\alpha \beta \gamma}(x_g) \ ,
\end{eqnarray}
where $O^{\alpha \beta \gamma}(x_g) \equiv O^{\alpha \beta \gamma}(x_1,x_2)$ with $x_g$ being the
total momentum transfer carried by gluons,
which can be conveniently chosen to be $x_g\equiv {\rm Max} \{ x_1,x_2  \}$.

\section{Summary}
In this paper, we have shown that an axial asymmetric color source distribution in the
transverse plane of transversely polarized proton can give rise to a spin dependent odderon in the MV model.
Such polarized odderon is responsible for SSA in jet production in the backward region of pp collisions
and SSA in open charm production in SIDIS process.
As a result, the BLV odderon solution can be tested by studying the x dependence of SSAs.
A relation between $k_\perp$ momentum of the odderon and the collinear twist-3 C-odd tri-gluon correlation
 has been established.
It is straightforward to extend our formalism  to study
SSAs  in open charm production in pp collisions and in Drell-Yan/direct photon processes,
 which already have been calculated in the collinear twist-3 framework~\cite{Koike:2011mb,Koike:2011nx}
 (for earlier work, see~\cite{Kang:2008ih}).
It is  our plan to further explore the possible difference/relation
between the CGC formalism and the collinear twist-3 approach in computing SSAs at small x.

\

\noindent
{\bf Acknowledgments:} I am grateful to Vladimir
Braun for bringing my attention to the potential relation between the odderon and the C-odd tri-gluon correlation.
This work has been supported by BMBF (05P12WRFTE).

\end {document}